# Platform Autonomous Custom Scalable Service using Service Oriented Cloud Computing Architecture

[1] B. Kamala [2] B. Priya [3] J. M. Nandhini

[1] AP-II, MCA Dept, Sri Sai Ram Engineering College, Chennai, kamala.mca@sairam.edu.in
[2] AP-II, MCA Dept, Sri Sai Ram Engineering College, Chennai, priya.mca@sairam.edu.in
[3] AP-II, MCA Dept, Sri Sai Ram Engineering College, Chennai, nandhini.mca@sairam.edu.in

**ABSTRACT**

The global economic recession and the shrinking budget of IT projects have led to the need of development of integrated information systems at a lower cost. Today, the emerging phenomenon of cloud computing aims at transforming the traditional way of computing, by providing both software applications and hardware resources as a service. With the rapid evolution of Information Communication Technology (ICT) governments, organizations and businesses are looking for solutions to improve their services and integrate their IT infrastructures. In recent years, advanced technologies such as SOA and Cloud computing have been evolved to address integration problems. The Cloud's enormous capacity with comparable low cost makes it an ideal platform for SOA deployment. This paper deals with the combined approach of Cloud and Service Oriented Architecture along with a Case Study and a review.

# 1. INTRODUCTION

Cloud computing refers to both applications delivered as services over the internet and the hardware and systems software in the datacenters that provide those services [1]. Cloud computing is represented by an "elastic" and "malleable" IT fabric or a platform which is exposed over the internet and enables enterprises to use their resources and services to the maximum potential. Cloud allows businesses to expand or contract the IT footprint, based on existent demand. It also provides a platform for developers to build customized applications. In cloud computing, infrastructure components are provided as hardware elements, software are provided as web services, applications (and toolkits) are exposed as APIs to the outer world. These elements, in turn, leverage and build a complete or partial IT landscape on cloud. Over the existing hardware, software, and application layers there is a nuance of management, governance, and processes.

## 1.1 Cloud – Advantages

The use of cloud computing has significant advantages such as
1. Cost reduction,
2. Great storage capacity
3. Scalability,
4. Needless software installation and maintenance
5. Accessibility of on-demand services or applications from anywhere,
6. Elasticity and
7. Pay-as-you-go model and energy saving.

Figure 1.1 illustrates the Cloud Computing topology [1].

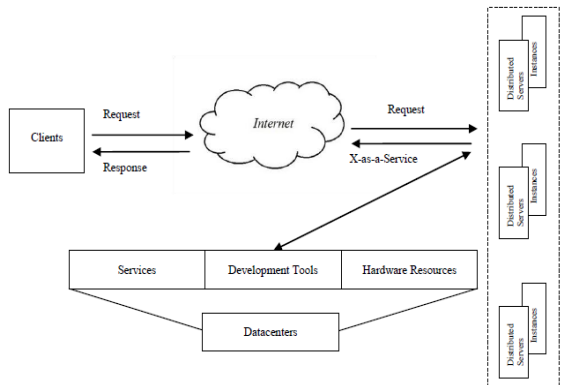

Figure1.1 Cloud Computing Topology

SOA is a design pattern which is composed of loosely coupled, discoverable, reusable, inter-operable platform agnostic services in which each of these services follow a well-defined standard. Each of these services can be bound or unbound at any time and as needed.

SOA attempts to streamline integration across systems by providing components that are architected and described in a consistent fashion. Web Services are required to implement SOA. Web Services technologies are based on providing common protocols with which clients can discover and contact the services through the World Wide Web.

Figure 1.2 depicts the overlap between Cloud and SOA.

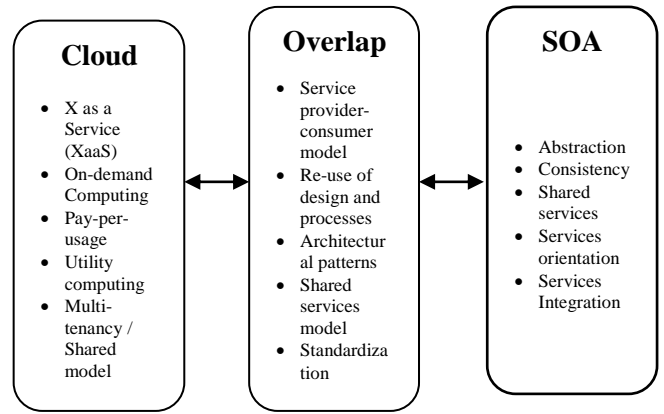

Figure 1.2 Cloud-SOA Overlap

The remainder of this paper is organized as follows: Section 2 deals with the new services in Cloud and the importance of SOA. Section 3 discusses the combined approach of Cloud and SOA. Section 4 discusses the Service-Oriented Cloud Computing Architecture. Section 5 deals with a case study on Health Care Domain.

## 2.0 NEW SERVICES IN CLOUD

Cloud computing includes Service Oriented Architecture (SOA) and virtual applications which is based on both hardware and software. This technology allows efficient computing by centralizing storage, memory, processing, bandwidth, etc. without knowledge of the user [2].

Software as a Service (SaaS) focuses on software hosted as a service while SOA focuses on software designed as a service.

**2.1 Benefits of SOA**

SOA offers positive benefits such as
1. Language – neutral integration
2. Component Reuse
3. Organizational agility
4. Leveraging Existing Systems

Cloud expands SOA by adding scalability and grid computing. Scalability is required when software is used as service due to the fact that as more and more of the service is instantiated, hardware resources are utilized and scalability becomes a requirement.

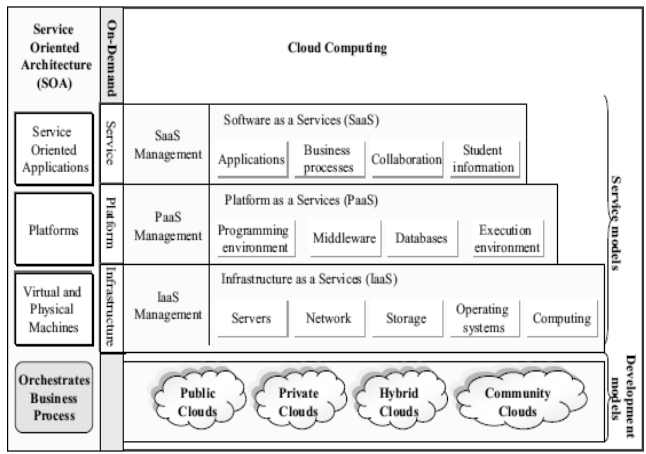

Figure 2.1 Relationship between SOA and Cloud Computing

Figure 2.1 shows the relationship between SOA and Cloud Computing.

**3.0 AN APPROACH ON MIX BETWEEN SOA AND CLOUD**

The various limitations concerning Cloud are:
1. Availability of services,
2. data lock-in risk,
3. Reliability of data,
4. Data transfer bottlenecks,
5. Performance unpredictability and software licensing are raised.

The adoption of Cloud and SOA results in a service full cloud computing environment that enables highly dynamic and effective organizational collaborations. In such collaboration each of the participants behaves according to the predefined and mutually agreed upon business logic and service level agreement.

A service-oriented architecture (SOA) may bring many benefits, such as: promoting reuse, ability to combine services for the creation of new composite applications, use of decoupled services through standard interface, offering at the same time a technological behavior for the development of business solutions. Implementation of Master Data Management in SOA strategy ensures data consistency, adjustment of information resources of the organization, correct dissemination of information in the interior/exterior of the university, as well as delivery of all potential benefits by SOA initiatives.

The objective of SOA is to create agility, fast reconfiguration, and new end to end processes and governance encompasses SOA at the strategic, tactical and operational levels. However, simply using a service-oriented architecture does not ensure organizational success [4]. Further, SOA brings additional levels of complexity because it focuses on both business processes and IT based services. Therefore, SOA should be used in combination with other tools or solutions in order to achieve organizational objectives.

The success of a cloud-model depends on its ability to multi-tenant combined with the necessary information security. In all of this, there is an underlying realm of re-use and service orientation that runs throughout; and it is at this realm where SOA supports cloud at the first instance.

SOA defines architectural principles for enterprise systems by defining interfaces, processes, and communication between various sub-systems, by focusing on predictable patterns and service behaviors. SOA encompasses a library or a repository of service components and processes that a service consumer can invoke. At a higher-level of encapsulation, SOA is a set of services collectively termed as web services, which are actually standards such as SOAP, WSDL, and so on. SOA advocates the principles of component reuse and a well-defined relationship between a service-provider and service-consumer. SOA enables the recurring use of existing application functionality and attempts to share common information, services, and processes across the organization.

In multiple ways, cloud can be seen as an extension of SOA that goes beyond applications and into IT infrastructure. IT components are increasingly being delivered as services, which has become the focus for IaaS and PaaS. This is truly a derivative of the SOA services. IT infrastructure components, similar to their software counterparts, have become to be discovered, managed and governed. Cloud computing can also viewed as the "cornerstone" that SOA was long awaiting for, that could see an increased adoption of SOA in the industry.

Table 3.1 lists out a few perspectives advocated by SOA that can be implemented in a cloud.

| S. No. | Perspectives in SOA | Implementation in Cloud |
|---|---|---|
| 1 | Standardization and Reuse of components | Virtual machine instances are built using standard template and provisioned for use. The template can be reused irrespective of the OS or the virtualization software |
| 2 | The ability to monitor points of information or service, in order to determine the health of the infrastructure. | Cloud implementations provide component, service, process-level monitoring of the IT footprint, through management tools. Also such tools provide clouds with the necessary service management and orchestration capability. |
| 3 | Architectural Discipline | Cloud environment for its success, implements a well defined and modular architecture at all levels – Infrastructure, applications and so on. |

Table 3.1 SOA Perspectives – Cloud Implementation

## 4.0 SERVICE ORIENTED CLOUD COMPUTING ARCHITECTURE:

The layered architecture of the Service Oriented Cloud Computing Architecture (SOCCA) is illustrated in the Figure 4.1[3]

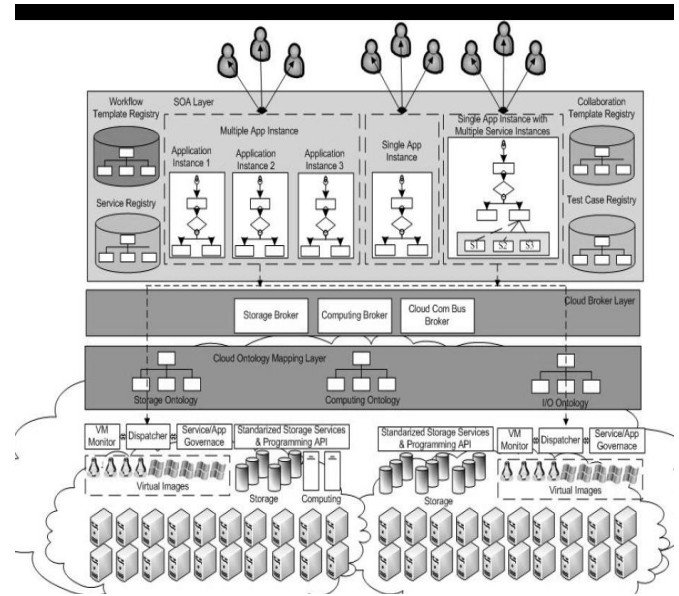

Figure 4.1 Service Oriented Cloud Computing Architecture

The Individual Cloud Provider Layer resembles the current cloud implementations. Each cloud provider builds its own data centers that power the cloud services it provides. Within each individual cloud, there is a request dispatcher working with Virtual Machine Monitor and Service/ governance service to allocate the requests to the available resources. The distinction from current cloud implementations is that the cloud computing resources in SOCCA are componentized in to independent services such as Storage Service, Computing Service and Communication Service with open-standardized interfaces, so that it can be combined with services from other cloud providers to build a cross-platform virtual computer on the clouds.

The Cloud Ontology Mapping Layer exists to mask the differences among the different individual cloud providers and it can help the migration of cloud application from one cloud to another.

Cloud brokers serve as the agents between individual cloud providers and SOA layer.

Each major cloud service has an associated service broker type.

The SOA layer fully takes the advantages of the existing research and infrastructure from the traditional SOA. The registry of each type of artifacts is indexed and organized by its ontology. The fundamental difference of the SOA layer of the SOCCA from the traditional SOA is that the service providers no longer host the published services. Instead, they publish the services in deployable packages which can be easily replicated and redeployed to different cloud hosting environments as shown in the Figure 4.2[3]

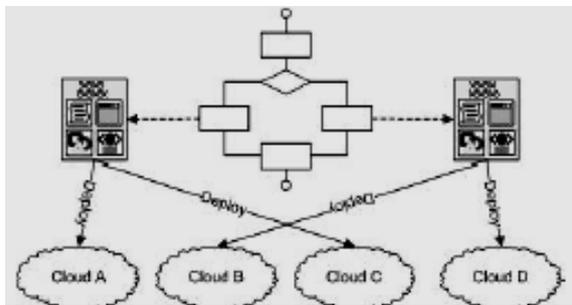

Figure 4.2 SOCCA Application Architecture

A workflow template is composed of service stubs/specifications, which specify the functionalities and interfaces of services. A service stub is then bound with a service package.

## 5.0 CASE STUDY – HEALTH CARE DOMAIN

The recent trend in today's IT health care system is to have single enterprise software that caters to all functional areas of the domain in an integrated and comprehensive manner. To achieve this objective various architectures have been implemented to develop an Enterprise Management System for health care domain. The latest trend in the technology leans towards SOA, Cloud Computing and integration using Universal standard modalities.

The various challenges that are faced by health care domain to achieve enterprise software are

- Each disparate system has its own standards.
- The clinical codification is different in different applications.
- Text analytics is a challenge.

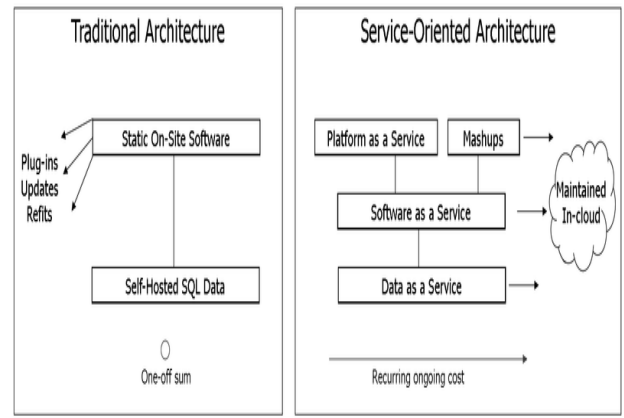

Figure 5.1 Traditional Vs SOA in Health Care Domain

As shown in the Figure 5.1[5], the SOA based architecture is designed to aggregate patient information into a comprehensive data repository, and access that information in near real time to improve decision making, which is a critical differentiator for healthcare organizations. This SOA based architecture facilitates building solution architectures that store and query documents from third-party components.

There are many benefits that result from Cloud technology in a healthcare setting. Some of the benefits are
- The ability to quickly access computing and storage resources when needed, without the requirement for a large technical staff.
- Improve customer services and operational efficiency at the same time.
- The range of services offered through the cloud is expanding at a staggering rate.

However, the complexity of cloud models, and the difficulty of measuring those risks, means that the stakes are far higher for sensitive information moving out into the cloud. The primary risks are, of course, going to be the availability, integrity and confidentiality of information. But the benefit of cloud models is that they hide much of the internal processes required to deliver the services.

The cloud also offers healthcare providers a unique opportunity to provide new services that would otherwise be cost prohibitive. Digital pathology is a good example. The move to digital pathology poses a huge problem for many healthcare providers as the costs associated with large file sizes and the required infrastructure is challenging to say the least. But a cloud based solution opens new doors. Not only could a facility manage the infrastructure requirements but they can provide access to specialists and pathologists that they may not have had access to before. This enables smaller facilities or remote caregivers to provide new services to their patient population in a cost effective manner.

Compared to centralized healthcare projects, the decentralized SOA-based solution has not only proven itself to be a successful approach, but has also resulted in cost reductions and improvements in the quality of healthcare all without compromising patient record privacy [6]. The centralized systems will usually have the edge in terms of scalability and accessibility; this decentralized approach has various capabilities for better security, greater cost savings and a faster implementation time.

## 6.0 CONCLUSION

This paper presents a review of need of new services in cloud along with the importance of SOA. SOA and cloud computing will coexists, complement and support each other. The SOCCA promotes an open platform on which open standards, ontology are embraced. A case study on the combined approach of SOA and cloud has been addressed in the Health care domain. As a result, SOA has helped Cloud to become what it is today and will continue to play a significant role in future. Increasingly, SOA is being connected well with cloud and there are ample applications of SOA principles in architecting the cloudscape. An enterprise should evaluate all aspects by comprehensively considering the pros and cons, before moving to cloud. Also a study of the implementation effects of the integrated solution (SOA and Cloud) can be carried out.